# An Intelligent Location Management approaches in GSM Mobile Network

N. Mallikharjuna Rao[1]
Department Of MCA,
APGCCS,
Rajampet, India
E-mail: drmallik2009@gmail.com

Prof M. M Naidu[2]
Department of CSE,
S.V College of Engineering,
Tirupati, India
e-mail: mmnaidu@gmail.com

P.Seetharam[3]
Systems Engineer
AITS
Rajampet
e-mail: seetharam.p@gmail.com

*Abstract -* **Location management refers to the problem of updating and searching the current location of mobile nodes in a wireless network. To make it efficient, the sum of update costs of location database must be minimized. Previous work relying on fixed location databases is unable to fully exploit the knowledge of user mobility patterns in the system so as to achieve this minimization. The study presents an intelligent location management approach which has interacts between intelligent information system and knowledge-base technologies, so we can dynamically change the user patterns and reduce the transition between the VLR and HLR. The study provides algorithms are ability to handle location registration and call delivery.**

## 1. INTRODUCTION

Recent advances in communication technology have created the opportunity for mobile terminals to receive many services that were, until not long ago, only available to tethered terminals. This system to support large scale mobility was the advanced mobile phone system. A new digital system, personal Communication System (PCS) provides voice as well as data services to wireless users. PCS works in the GSM 800/1900 MHz spectrum. There are competitive standards for analog, digital, and PCS system throughout the world [1].

One of the challenging tasks in a PCS environment is to efficiently maintain the location of the PCS subscribers in GSM who move around freely with their wireless unit. In India TRAI (Telephonic Regulatory Authority of India) is used for managing location information of the subscribers and enabling them to send and receive calls and other services such as messaging and data service.

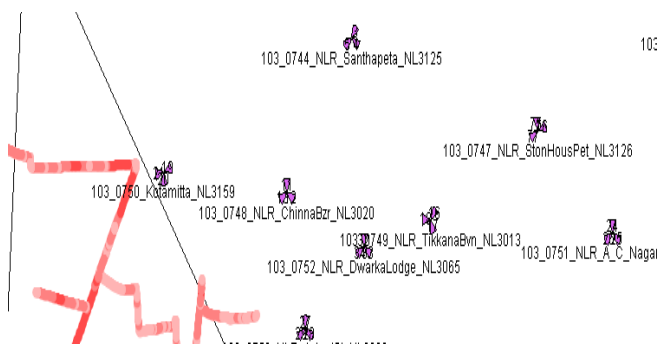

Figure 1: Sample Location Areas

The cells are established as we have shown in figure 1, using which mobile subscribers move between them and he make calls for transmitting voice/data.

The network reference model of a Personal Communication System in GSM network is shown in figure 2 we refined for supporting and understanding of my work .It consists of the following components [2].

*Home Location Register* (HLR): Maintains the profiles of the entire subscribers that are registered with the home network. When a mobile subscriber roams to another area, it has to register with the Visitor Location Register (VLR) of that area. The HLR maintains a pointer to the VLR which currently serve the mobile.

*Visitor Location Register* (VLR): Supports registration, authentication, and call routing to/from a mobile while it is away from its home area. Each MSC has a VLR to holds the data relevant for handling calls from and to the MSs that are currently located in its area. The relevant data is downloaded from the home HLR when the mobile subscriber switches on the mobile handset in the area of the visited MSC thereby initiating the process of registration. VLR holds the exact location of the MS and keeps on updating the location as the mobile move across its area.

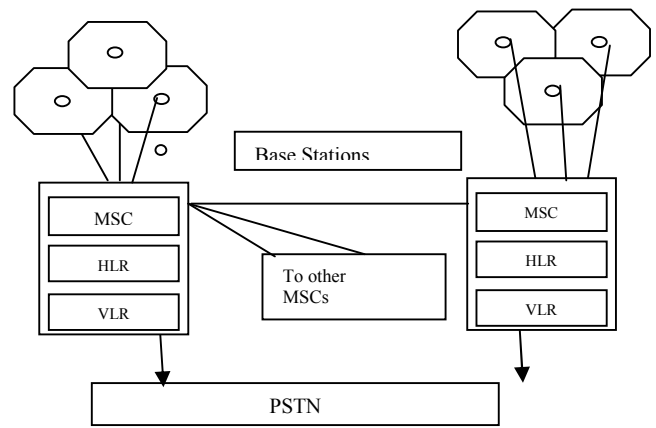

Figure 1: Network Reference Model

*Mobile Switching Center* (MSC): Responsible for switching the voice/data connection to the mobile host. GMSC is Gate way Mobile Switching center which can route the calls from PSTN.

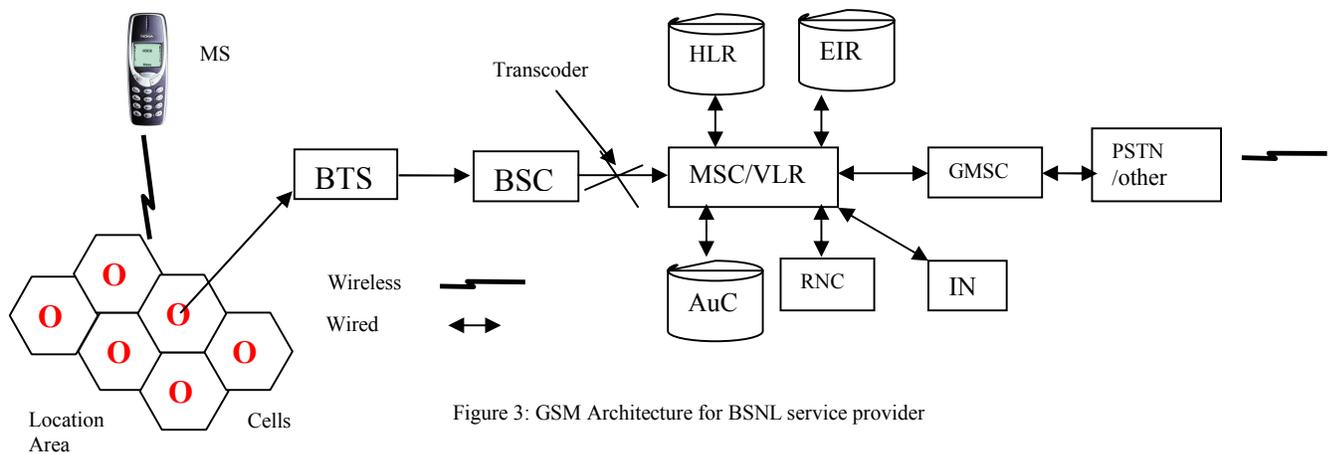

Figure 3: GSM Architecture for BSNL service provider

*Base Station* (BS): The base station is the gateway between the wireless network and wired network. It provides the wireless connection to the mobile subscribers within its coverage area (Cell). A set of base stations are connected to the MSC through a Base Station controller.

Authenticating Center (AC or AuC): The Authentication Center is a workstation system, which authenticates subscribers. AuC needs to access user information for authentication process so it is co-located with HLR.

Equipment Identify Register (EIR): It is a database which stores information for the identification of mobile units.

Public Switched Telephone Network (PSTN): This component refers to the regular wired line telecommunication network which is commonly accessed by landline calls.

Integrated Service Digital Network (ISDN): It is a wired line network which provides enhanced digital services to subscribers.

For supporting our work, we studied Nellore District BSNL office functionality and we find the existing architecture for GSM network as shown in figure 3.

Every subscriber is registered with a home network, the HLR of which maintains the subscriber's current physical location. This physical location is the LAI is the ID of the MSC currently serving the subscriber. If the subscriber has moved to another region then he/she has to register with the VLR that covers the new region. During registration, the VLR will contact the subscriber's HLR, and the HLR will update its database to reflect the new location of the subscriber. If the mobile has registered with some other VLR before, HLR will send a registration cancellation message to it.

Present system is having the disadvantage in call setup and call delivery using the traditional centralized database system. One disadvantage is that since every location request as well as location registration are serviced through a HLR, in addition to the HLR being over loaded with database lookup operations [1] [3] [6]; the traffic on the links leading to the HLR is heavy. The other disadvantage is that any HLR to be unreachable even though mobiles may be roaming and away.

In view of above disadvantages, we are proposing intelligent location management approaches for to overcome all the disadvantages faced by the system. This effort will reduce the transition time between the VLR and HLR databases when the common MS roams to that MSC services area

In this paper, fuzzy logic concept is used for supporting proposed intelligent location management schemes, fuzzy logic deals with vague, doubtful and ambiguous data for giving better results for uncertain data in mobile networks. Vagueness or doubtfulness means that cannot be defined or determined data. In general, vagueness is as associated with the difficulty of making sharp or precise distinctions in the world. That is some domain of interest is vague if it cannot be delimited by sharp boundaries [14].

Ambiguity is associated with one-to-many relations that are situations in which the choice between two or three more alternatives is left unspecified. In this is paper we are proposing the techniques that can work easily to see that the concept of a fuzzy sets provides a base mathematical framework for dealing with vagueness.  Section 2 describes the existing databases system and approaches, section 3 describes the proposed intelligent system and concluding remarks is presented in section 4.

## II. BACKGROUND AND EXISTING SYSTEM

Location management includes two major tasks: Location registration and call delivery. Location registration procedures update the location database (HLR and VLRs) when an MT moves into different location area. Call delivery procedure locate a Mobile Station (MS) based on the information available at HLR and VLRs when a call for the MS is initiated.

In GSM network, there are two kinds of databases: HLR and VLR are used to store the location information of MSs. The whole network coverage area is divided into cells as shown in figure 1. There is a base station installed in each cell and an MS within a cell communicates with the network through a base station. These cells are grouped together to from a larger area called a registration area (RA)/Location

Area (LA). All the base stations belonging to one LA are wired to a mobile switching center (MSC) through BTS and BSC which serves as the interface between the wireless and the wired networks. In this paper, we assume that one VLR is associated with each MSC as we shown in figure3.

A Location Area (LA) as shown in figure 3 is defined as a group of cells. Within the network, a subscriber location is known by the location area which they are in and is controlled by a Base Station Controller (BSC). The identification of a location area in which an MS is currently located is stored in the VLR.

When a MS crosses a boundary from a cell belonging to one LA into a cell belonging to another LA, it must report its new location to the network. When a MS crosses a cell boundary within a location area, it does need to report its new location to the network. But when a MS leaves the MSC service area, the scenario changes and updating information is increased. The VLR is always integrated with the MSC as shown in figure 3 and there is one VLR for each MSC service area.

The VLR can be regarded as a distributed HLR as it holds a copy of the HLR information stored about the subscriber [5]. The data stored includes:

1. Mobile subscriber Roaming number (MSRN)

2. Service type (services that the subscriber is allowed to access)

3. Current location

4. HLR address Ciphering keys

5. Billing information

6. International Mobile Subscriber Identity (IMSI)

7. Subscribers phone number

8. Access point subscribed (GPRS)

9. Temporary Mobile Subscriber Identity (TMSI)

When a subscriber roams into a new MSC service area, the following steps occur in figure 4 [5]

*Step1*: The VLR checks its database to determine whether or not it has a record for the MS. And this checking is based on the subscribers IMSI.

*Step2*: VLR sends a request to the subscribers HLR for a copy of the MS's subscription when it does not find any record for the corresponding MS

*Step3*: the HLR passes the information to the VLR and updates its location information for the subscriber.

*Step4*: the HLR instructs the old VLR to delete the information that was stored in the database.

*Step5*: the VLR stores its subscription information for MS, including the latest location and status (idle). This maintains of two databases at HLR and VLR gives a flexible mechanism to support call routing and dialing in a roaming situation.

Two major steps are involved in call delivery: determining the serving VLR of called MT, and locating the visiting cell of called MT. The following step occurs in call delivery shown in figure 5 [5]

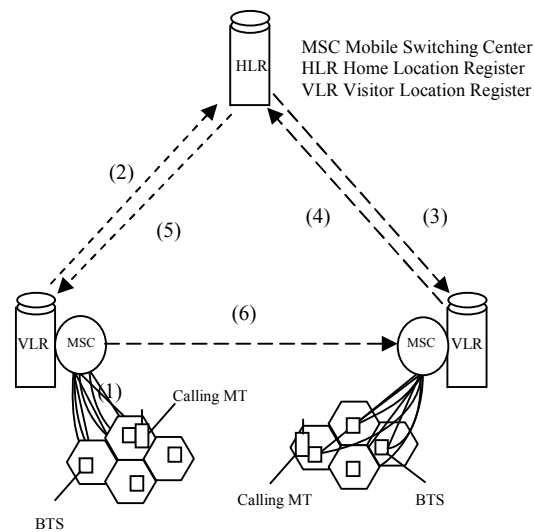

Figure 5: Call delivery procedure

*Step1*: Calling MT sends call initiated signal to its serving MSC through the base station.

*Step2*: MSC of calling MT, calling MSC, sends the location request message to HLR of called MT.

*Step3*: HLR of called MT determines the current serving VLR of called MT and sends the route request message to the associated MSC, called MSC.

*Step4*: Called MSC determines the cell location of called MT and assigns temporary location directory number (TLDN) to called MT. called MT MSC then sends this routing information (TLDN) to HLR

*Step5*: HLR forwards TLDN to calling MSC

*Step6*: calling MSC requests a call setup to called MSC through the SS7 network.

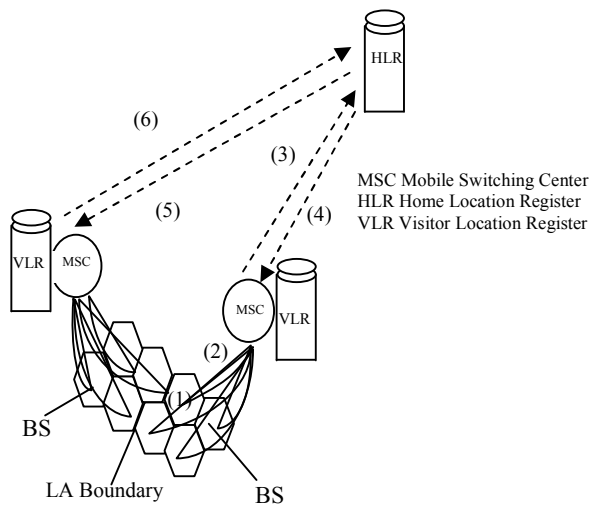

Figure 4: Call registration procedure

In the existing system, when subscriber is moved from one base station to other base station, subscriber's details are transferred to another base station VLR (data is transferred from HLR). If call is disconnected the same record stored at VLR is deleted automatically. For solving above problem we are proposing new approaches and these approaches are motivated from several works which were already done. In past studies [6] they proposed a novel mechanism for the VLR interaction methods where it has been given the potential to be intelligent to find the common users for the MSC service area. This effort will reduce traffic load and also improve spectrum efficiency. Moreover they concluded that, it reduces transition between VLR and HLR Therefore, it can play a significant role to support large number of traffic in the future cellular networks. In work [7] similar to [6] but they proposed fVLR, registration procedure of a mobile station location is described and a call setup procedure. They said to store and manage subscriber visits in fVLR database of which a mobile user frequently visits location. And then when set up the call path between mobile users, the VLR of the caller queries callee's fVLR for searching the location of callee instead of requesting to HLR of the callee. In this work fVLR can be well applied to the mobile users that live a well regulated life. In another study [10] proposed intelligent approach by taking the User Profile History (UPH); to reduce the location update cost. They obtained results the efficiency of UPH in significantly reducing the costs of both location updates and call delivery. In another study [12] proposed fuzzy logic approaches for user location tracking which is highly motivated us to work on these approaches for location management. In this study [13] demonstrate the sum of update and lookup costs of the location database must be minimized. Along with these

As per intelligent database system for location management, record will not be removed immediately from the VLR even subscribers is changed his/her MS. Proposed system will store for few days if any repetitions calls to the same base station occurs. Its means that system will act intelligently to take the decision, then removes the record from the VLR.

## 3. PROPOSED SYSTEM

Usually when the Mobile Station leaves one MSC service area to another, all the information related to the Mobile Station user is deleted (based on the subscribers IMSI or TMSI). Here in the proposed method deletion will be done depending on a decision that will be taken after doing some analysis during certain amount of time frame: 1) The VLR will have been observing the mobile station roaming habit for some days , for example 'n' days. 2) The VLR will also observe the time when the Mobile station roams into the MSC area. Depending on these data, a query will be done and the result of the query will help to decide that whether the information of mobile station will be stored inside the VLR or be dropped. The goal of the query is to find those mobile users who used to visit the MSC service area at least once regularly during the observation period. The found regularly visiting Mobile station will be termed as common mobile station. Some information like account balance, service validity and service that subscriber is allowed to access, is always needed to be updated.

### A. Proposed Model Scenario

Let us consider a Mobile Station user travels 115 km daily. He lives at 'Nellore' and goes to his office at 'Rajampet' which is 115 km apart from his house. Since he is to drop his wife and daughter at 'Podalakur' and 'Rapur' respectively, he is to travel same path daily. Now the $VLR_1$ will find that the person leaves this area at 8 am and returns at 8 pm. And the $VLR_2$ of MSC service area will find that the person enters this service area at 8.30 am and leaves at 8.40 am daily. Similarly we assume, the possible roaming time in different MSC service area (here in this kadapa area is come $MSC_1/VLR_1$ and Rajampet comes in $MSC_2/VLR_2$ area). The same information is used for call delivery strategy.

In this paper, we are proposing intelligent location management scheme for finding the frequently visited mobile subscribers in particular location area, as we said above use fuzzy logic, the fuzzy sets within the field of decision making have for the most part consisted for extensions or "fuzzifications" of the classical theories of decision making.

Classical decision making generally deals with a set of alternatives comprising the decision space, a set of states of nature comprising the state space, a relation indicating the state or outcome to be expected from each alternative action.

### B. Decision making of intelligent database VLR using Fuzzy Logic

Now we consider during the '$m$' days of observation period, the number of roaming users is '$N$' on all (week) working days respectively in Mobile Switching Center service area. Then, the observation set for each day can be considered as follows. For example, a mobile subscriber roams from $LA_1$ to $LA_2$ frequently, intelligent system process the following:

Subscriber moves from $LA_1$ to $LA_2$, subscriber profile reads from HLR and sends to existing VLR for subscriber identity to make a call. The subscriber moves from existing $LA_2$ to $LA_3$ and then subscriber record will be deleted automatically and subscriber's identity will be send to $LA_3$. Suppose, if Subscriber again visits $LA_2$, subscriber profile reads from HLR and is send to VLR for identity, which we are presented in scenario model. But, here we have to remember one thing, that how many times the mobile subscriber visits or roams in one particular LA, this is called uncertainty. In this paper, we are presenting the fuzzy decision making for handling such uncertainties to retrieve subscribers records from VLR database. The fuzzy model of decision making is proposed by Belleman and Zadesh [1970] [11], which we are illustrating by a simple example. Suppose we choose different LA and their VLR sets which are mentioned below in mobile switching center:

Fuzzy Set= $\{(LA_1, VLR_1), (LA_2, VLR_2)… (LA_n, VLR_n)\}$

The characteristics function of a crisp set assigns a value of either 1 or 0 to each individual in the universal set, thereby discriminating members and nonmembers of the crisp set under consideration. This function can be generalized such that the value assigned to the elements of the universal set fall within a specific range and indicate the membership grade of these elements in the set in question. Larger values denote higher degree of set membership. Such a function is called a membership function and the set defined by it's a fuzzy set [11]. The range of values of membership functions is the unit interval [0, 1]. Here each membership function maps elements of a given universal set X, which is always a crisp set, into real numbers in [0,1]

The membership function of fuzzy set A is defined by a,

A: → X [0, 1]

Once a fuzzy decision has been arrived at, it may be necessary to choose the "best" single crisp alternative from this fuzzy set. A fuzzy set may be represented by a meaningful fuzzy label. A reasonable expression of these concepts by trapezoidal membership $S_1$, $S_2$, $S_3$ these functions are defined on the interval [0, 20]. For example, "Low_visits", "Medium_Visits", "High_Visits" are linguistic variable for fuzzy set frequently visited locations maintained in mobile VLR databases. The linguistic representation is as follows:

$$\text{Low\_Visits} = \begin{cases} 1, & \text{No\_of\_visits} \leq 4 \\ (8-\text{No\_of\_visits})/5, & 4 < \text{No\_of\_visits} < 8 \\ 0, & \text{No\_of\_visits} \geq 8 \end{cases}$$

$$\text{Medium\_Visits} = \begin{cases} 0, & \text{No\_of\_visits} \leq 8 \\ (\text{No\_of\_visits}-12)/5, & 9 < \text{No\_of\_visits} < 12 \\ (14-\text{No\_of\_visits})/5, & 12 \leq \text{No\_of\_visits} < 14 \\ 1, & 14 < \text{No\_of\_visits} \geq 15 \end{cases}$$

$$\text{High\_Visits} = \begin{cases} 0, & \text{No\_of\_visits} < 16 \\ (18-\text{No\_of\_visits})/5, & 16 \leq \text{No\_of\_visits} < 18 \\ 1, & \text{No\_of\_visits} \geq 18 \end{cases}$$

Figure 6: Triangular membership functions for Visits

Suppose, if we assume three mobile subscribers they are visiting and registered at $VLR_1$ database.

Mobile-subscribers set = {($S_1$, 5), ($S_2$, 12), ($S_3$, 17)} as we shown in table 1.

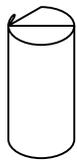

| ID | Common Visits | Current visits |
|---|---|---|
| $S_1$ | 3,5,6 | 5 |
| $S_2$ | 2,3,4 | 12 |
| $S_3$ | 3,4,5 | 17 |

Table 1: Visitors info in VLR database

In this above table, if we observe $S_1$ has visited 5 times on the days of 3, 5, and 6 in a 7 day format. That means he/she has visited more than one time in day. The decision making scenario using fuzzy logic is, so we can calculate maximum number of times visited to $VLR_1$ in $LA_1$, the membership degree is:

Subscriber $S_1$ visited to $VLR_1$ 5 times

Subscriber $S_2$ visited to $VLR_1$ 12 times

Subscribers $S_3$ Visited to $VLR_1$ 17 times

Membership degree for subscriber $S_1$ is 8-5/5 = 3/5 =0.6

Membership degree of subscriber $S_2$ is 14-12/5= 2/5=0.4

Membership degree of subscriber $S_3$ is 18-17/5=1/5=0.2

The intersection for fuzzy set is

$\mu_{A \cap B}(x) = \text{Min}[\mu_A(x), \mu_B(x)]$

Therefore, to find the frequently visited subscriber is

= Min [$S_1$/0.6, $S_2$/0.4, $S_3$/0.2] or Min [0.6, 0.4, 0.2]

= $S_3$ (0.2)

= Frequently visited Subscriber is $S_3$

This fuzziness allows the decision maker to frame the goals and constraints in vague, linguistic terms; which may more accurately reflect the actual state of knowledge. In this model, we have shown to stored the linguistic values in fuzzy VLR in figure 7.

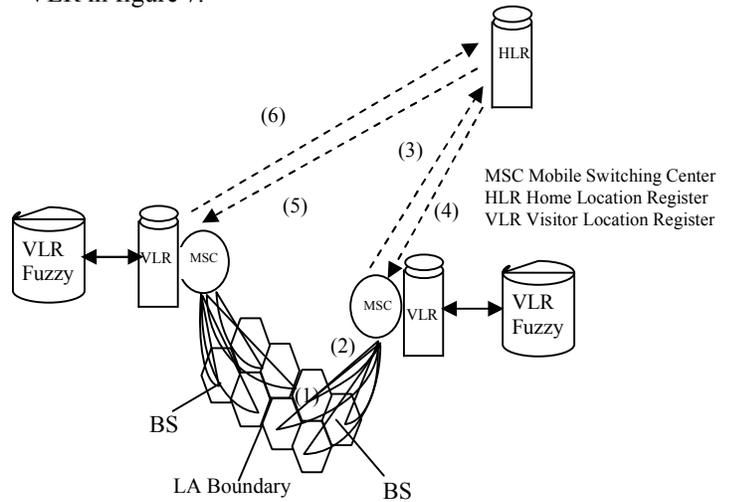

Figure 7: Proposed Intelligent Call registration

As we shown in figure 7 VLR database is divided into two tiers one is low-tier and second one is high-tier DBs. They are represented as V (Traditional VLR) and $V_1$ (Fuzzy VLR). When the subscribers come into LA, MS functions checks in V and V will check at $V_1$, if it found that there is no need to search the subscriber's profiles from the Home Location Register (HLR). i.e. there are two operations are performed in this system. They are VLR traditional (tier1) → VLR fuzzy (tier2). Immediately VLR fuzzy (tier 2) → VLR traditional (tier1) transfers the old record to VLR. In this scheme, VLR fuzzy is stores the information if subscribers are frequently visited subscribers in that location area. Record will be deleted after specified period automatically from both the VLRs. In this case, the fuzzy constraints may be defined on the set *V* and fuzzy goals on the set $V_1$ such that

$\mu_C: V \to [0, 1]$ and $\mu_G: V_1 \to [0, 1]$

function *f* can then the defined as a mapping from the ser of actions *V* to the set of outcomes $V_1$ *f: V → $V_1$* such that a fuzzy goal G defined on set V induces a corresponding fuzzy

$$\mu_D(x) = \text{Min}\,[\mu_G(x), \mu_C(x)]$$

goal G' on the set V, thus $\mu_G'(x) = \mu_C(x)$ a fuzzy decision D may them be defined as the choice that satisfies both the goals G and the Constraints C. if we interpret this as a logical, we can model it with the intersection of the fuzzy sets G and C D= G∩C the fuzzy decision D is the specified by the membership function

*C. Algorithm for Intelligent VLR*

The Location management algorithm makes use of the fact that an average mobile user has limited number of frequently visited locations. Usually a type of activity such as work, school, or shopping, occurring at a particular location. As an example, the shopping locations are located at specific sites, and for most users, the work, school, and home locations are fixed. We are considering 'm' days of observation period for a VLR to find its common mobile station and the searching must be based on IMSI of the visited MS during that time frame. A possible algorithm is given below:

*Integer* get_common_MS ()

```
{       Integer i=0; k=0; m=0;
     Interger array AA[m][];
    AA[m] AA[m] = days_IMSI_copy(1,2,3,4,5,6,7);
   b_day = get_busiest_day_user();
 Max=max1=maximum_user_in_busy_day ();
    // comparing to comparing to get common IMSI;
   for ( ; i<=max; i is incremented by 1)
   {
     for (; j<=m; j is incremented by 1)
     for (;k<max1;k is incremented by 1)
       {
            If AA[j][k]=b_day[i])
            {  d is incremented by 1
            If (d==m) copy _IMSI (A[i]);
            }
     } } }
```

Figure 6: Proposed Intelligent VLR algorithms

*D. Algorithm for Intelligent call delivery using fuzzy Logic*

Frequently visited locations of a user are not pre-defined in location management, since this information is not always collected from mobile subscriber. whenever a subscriber want to make a call, profile identification is required to check whether the subscriber identity is verified and accepted by HLR or not, but in our proposed system, if subscriber is a known visitor of that particular location, call will be registered in VLR and will not be deleted immediately by the VLR register after subscriber moves out of VLR, as we shown in figure 7, the visited records are saved in fuzzy VLR, we will fix up time to each and every subscriber for keeping his profiles in VLR database, if they do not visit that location with in the specified time, the subscriber record then be deleted from VLR as well as Fuzzy VLR. Therefore, it is the responsibility of the location management algorithm to intelligently determine such information.

The intelligent algorithm determines these frequently visited locations of a user at a given time of day based on the individual user profile.

*E. Proposed call delivery algorithm*

In this system, VLR maintains two entries for storing the data as we shown in figure 8. One is for the high-tier database is called VLR and another is for the low-tier database is called Fuzzy VLR for storing fuzzy related values. Mobile subscriber roams in the same area frequently; it not required in retrieving the subscriber record from HLR for identity that record will be there in fuzzy VLR database up to some period. If the subscriber will not visit same location in specified period of time record then the record will be deleted automatically. Call delivery setup will have two Databases one as HLR and another as VLR, in our proposed system the VLR database will be divided into two tier architecture database which will have High tier database as VLR and Low Tier database as VLR Fuzzy. Subscribers make a call data transfer in the following way:

VLR $\longrightarrow$ VLR$_{Fuzzy}$ if subscribers profile is available call will be delivering without asking HLR because subscriber is visits the same location frequently. If subscriber is new for that location that means VLR$_{fuzzy}$ send signal VLR and VLR ask subscriber profile from HLR. (i.e. VLR$_{fuzzy}$ $\longrightarrow$ VLR). These two tier system as we shown in figure 8 for call delivery process.

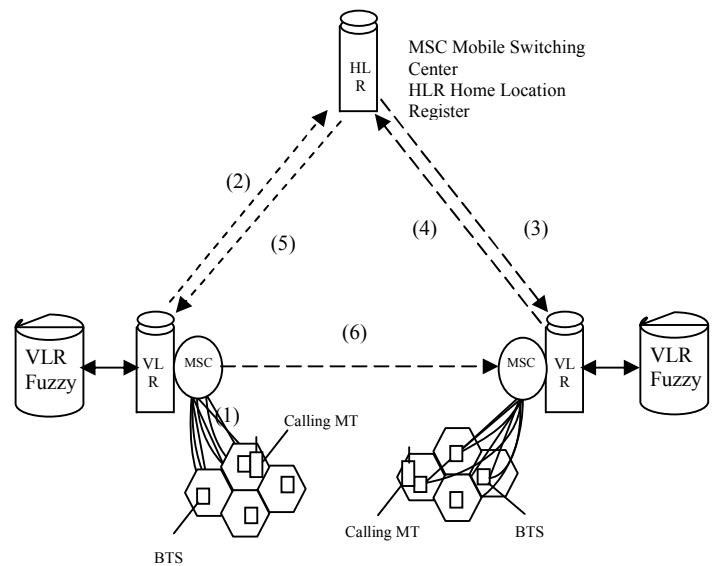

Figure 8: Proposed Intelligent Call delivery

We present trapezoidal membership for mobile subscribers, who are visiting frequently on the interval [0, 7]. For example, "Low_visits", "Medium_visits", "High_Visits" are linguistic variable for fuzzy set frequently visited locations maintained in mobile VLR databases. The linguistic representation is as follows for identify the subscribers are visiting the same location area or not.

By using fuzzy relations we tested for 7 days and stored it in and as we show the proposed fuzzy VLR database values in table 2. Intelligent call delivery algorithm is derived basing on the values shown in the Table 2.

$$\text{Low Visits} = \begin{cases} 1, & \text{No\_of\_visits} < 1 \\ (\text{No\_of\_visits}-1)/3, & 1 < \text{No\_of\_visits} \leq 2 \\ 0, & \text{No\_of\_visits} > 2 \end{cases}$$

$$\text{Medium\_Visits} = \begin{cases} 0, & \text{No\_of\_visits} > 2 \\ (\text{No\_of\_visits}-2)/3, & 2 < \text{No\_of\_visits} < 4 \\ (6-\text{No\_of\_visits})/3, & 4 \leq \text{No\_of\_visits} < 5 \\ 1, & \text{No\_of\_visits} < 5 \end{cases}$$

$$\text{High Visits} = \begin{cases} 0, & \text{No\_of\_visits} > 5 \\ (7-\text{No\_of\_visits})/3, & 5 \leq \text{No\_of\_visits} < 6 \\ 1, & \text{No\_of\_visits} < 6 \end{cases}$$

Figure 9: Triangular membership functions for visits in a week

| Mob ID | Visits in a week | | | | | | | No. of Visits | Freq. Visits | Time Expiry |
|---|---|---|---|---|---|---|---|---|---|---|
| | $D_1$ | $D_2$ | $D_3$ | $D_4$ | $D_5$ | $D_6$ | $D_7$ | | | |
| $S_1$ | 0 | 0 | 0 | 0 | 0 | 0 | 1 | 1 | Low | week |
| $S_2$ | 0 | 0 | 0 | 0 | 0 | 0 | 0 | 0 | Low | week |
| $S_3$ | 1 | 0 | 1 | 0 | 1 | 0 | 1 | 4 | Medium | week |
| $S_4$ | 1 | 1 | 0 | 1 | 1 | 1 | 1 | 16 | High | week |

Table 2: Subscribers visiting statistics

Call delivery setup and deletion of subscriber profile at VLR database will be depended upon the algorithm given through Fuzzy. For example $S_1$ Subscriber visits are limited to 1, $S_2$ Subscriber visits are limited 0, in case of $S_2$ Subscriber the VLR database will delete $S_2$ profile immediately after 7 Days (we have considered 7 days for observation). Likewise subscriber frequency of visits is defined as Low, Medium and High as fuzzy linguistic variables. A possible algorithm is given below:

```
Integer get common_Call ()
{
    Integer i=0; k=0; n=0;
    Integer array hlr[n][]
    ?? n =7
    ?? ans = ??;
    ?? Test =??;
    ?? i = ??;
    ?? k =??;
    ??[][] hlr = new??[??][??];
    Test = 1;
    While (test > n)
    {
        ans = get???("specify that a Subscriber is arrived or not (0/1)");
        hlr[i][k] = ans;
        k = k + 1;
        Test = test + 1;
    }
}
```

Figure 9: Proposed Intelligent Call Delivery algorithm

We developed above algorithms by using Raptor, Raptor is a simple-to-use problem solving tool that enables the user to generate executable flowcharts and algorithm.

## IV. RESULTS

We felt that we have the following Advantages with the proposed location management approaches:

- ✓ As we understand that the proposed schemes are reduce the transition between VLR and HLR database because search/update/delete operations of the subscriber profiles every time from the HLR database whenever the subscribers are reached to same location area or if it is already roamed by subscriber earlier. However, this proposed mechanism also reduces the transition between MS and MSC.

- ✓ Call delivery algorithm, which can reduce the call delivery latency in the intelligent registration scheme because subscriber record is already available at $VLR_{Fuzzy}$ database hierarchy. It is avoided multiple registrations for call delivery.

- ✓ As we said earlier, fuzzy databases are more flexible than traditional database to querying the data from the databases. Hence, the proposed approaches will reduce the retrieval cost also.

- ✓ Moreover, Fuzzy databases are support multi key file structures to retrieve the records very fast and accurately.

## V. CONCLUSIONS

Mobile subscriber moves any where in mobile networks, and then location registration is need always for making calls. In this study, we noticed how these two databases Home Location Register (HLR) and Visitors Location Register (VLR) reflects when subscribers make call. When subscriber makes a call two basic operations are performed. In this paper, we proposed intelligent location management schemes; they can reduce the cost of maintaining location of mobile users by using Fuzzy Logic and Fuzzy databases for call registration and call delivery procedure of a mobile station for describing location and a call setup. We demonstrate the how to store and manage fuzzy data crisp values of which a mobile user frequently visits the location. Our proposed fuzzy based VLR can be well applied to the mobile subscribers that live a well regulated life. In the future we will further research improved on mobility management using fuzzy logic and fuzzy databases.

AUTHOR PROFILE

N. Mallikharjuna Rao is presently working as Associate Professor in the Department of Master of Computer Applications at Annamacharya PG College of Computer Studies, Rajampet and having more than 12 years of Experience in Teaching UG and PG courses. He received his BSc (Computer Science) from Andhra University in 1995, Master of Computer Applications (MCA) from Acharya Nagarjuna University in 1998, Master of Philosophy in Computer Science from Madurai Kamarj University, Tamilnadu, in India and Master of Technology in Computer Science and Engineering from Allahabad Agricultural University, India. He is a life Member in ISTE and Member in IEEE, IACSIT, and IAENG. His research interests are on Mobile networks, Mobile databases and Fuzzy Databases. He is a research scholar from Acharya Nagarjuna University under the esteemed guidance of Dr. M. M. Naidu, Dean & Professor, Dept of CSE, SV University, Tirupati, AP.